%% file: ms.tex
\documentclass{article}
\usepackage{spconf,amsmath,graphicx,bbm}
\usepackage{times}
\usepackage{epsfig}
\usepackage{amssymb,color}
\usepackage{enumitem} 
\setitemize{noitemsep,topsep=0pt,parsep=0pt,partopsep=0pt}
\usepackage{subcaption}

\title{Aggregation and Embedding  for Group Membership Verification}
\name{\begin{tabular}{c}Marzieh Gheisari$^{\dagger}$, Teddy Furon$^{\dagger}$, Laurent Amsaleg$^{\dagger}$, \thanks{Research supported by the ERA-Net project ID\_IoT 20CH21\_167534.} 
		 Behrooz Razeghi$^{\star}$, Slava Voloshynovskiy$^{\star}$ \end{tabular}}
\address{$^{\dagger}$ Univ Rennes, Inria, CNRS, IRISA, France, $^{\star}$University of Geneva, Switzerland\\
	{\small \{ \texttt{marzieh.gheisari-khorasgani, teddy.furon}\}@\texttt{inria.fr}},
	{\small \texttt{laurent.amsaleg}@\texttt{irisa.fr}},\\
	{\small \{ \texttt{behrooz.razeghi, svolos}\}@\texttt{unige.ch}
	}
}
\begin{document}

\newcommand\vect[1]{\mathbf{#1}} 
\newcommand\func[1]{\mathsf{#1}} 
\def \group {\mathcal{S}}
\def \x {\vect{x}} 
\def \y {\vect{y}} 
\def \z {\vect{z}} 
\def \f {\vect{f}}
\def \X {\vect{X}}
\def \Y {\vect{Y}}
\def \Z {\vect{Z}}
\def \w {\vect{u}} 
\def \m {\vect{m}} 
\def \a {\vect{a}}
\def \n {\vect{n}} 
\def \W {\vect{W}} 
\def \G {\vect{G}} 
\def \real {\mathbb{R}} 
\def \dim {d} 
\def \rep {\vect{r}} 
\def \hyp {\mathcal{H}} 
\def \Pr {\mathbb{P}} 
\def \Pfn {P_{\mathsf{fn}}}
\def \Pfp {P_{\mathsf{fp}}}
\def \Ptp {P_{\mathsf{tp}}}
\def \pfn {p_{\mathsf{fn}}}
\def \ptp {p_{\mathsf{tp}}}
\def \pfp {p_{\mathsf{fp}}}
\def\un{{\mathbbm{1}}}
\def\E{{\mathbbm{E}}}
\def\Var{{\mathbbm{V}}}
\def\0{{\mathbf{0}}}
\def\etal{{\it et al.}}
\def\ie{{\it i.e.}}
\def\eg{{\it e.g.}}


\maketitle

\begin{abstract}

This paper proposes a group membership verification protocol preventing the curious but honest server from reconstructing the enrolled signatures and inferring the identity of querying clients. The protocol quantizes the signatures into discrete embeddings, making reconstruction difficult. It also aggregates multiple embeddings into representative values, impeding identification. Theoretical and experimental results show the trade-off between the security and the error rates.
\end{abstract}
\input{introduction}
\input{RelatedWorks}
\input{ProblemFormulation}
\input{Figure12}

\input{OneGroup_L}
\input{MGroups}

\vspace{-10pt}
\section{Conclusion}
\vspace{-9pt}
This paper proposed four schemes for verifying the group membership of continuous high dimensional vectors.  The keystones are the aggregation and embedding functions. They prevent accurate reconstruction of the enrolled signatures, while recognizing noisy version.  However, the anonymity is slightly revealed when managing many signatures aggregated into several representatives: the server is only able to link each signature to its group number. Yet, the full identity of the user is preserved.


\newpage

\bibliographystyle{IEEEbib}
\bibliography{ms}

\end{document}

%% file: introduction.tex

\vspace{-8pt}

\section{Introduction}
\label{sec:Introduction}
\vspace{-4pt}

Verifiying that an item/device/individual is a member of a group is needed for many applications granting or refusing access to sensitive resources. Group membership verification  is \emph{not} about identifying first and then checking membership. Rather, being granted with access requires that the members of the group could be distinguished from non-members, but it does not require to distinguish members from one another. 

Group membership verification protocols first enroll eligible \emph{signatures} into a data structure stored at a server. Then, at verification time,  the structure is queried by a client signature and the access is granted or not. For security, the data structure must be adequately protected so that a honest but curious server cannot reconstruct the signatures. For privacy, verification should proceed anonymously, not disclosing identities.

A client signature is a noisy version of the enrolled one, \eg\ due to changes in lighting conditions. The verification must absorb such variations and cope with the continuous nature of signatures. They must be such that it is unlikely that a noisy version for one user gets similar enough to the enrolled signature of any other user. Continuity, discriminability and statistical independence are inherent properties of signatures.

\begin{figure}
\centering
\includegraphics[scale=0.435]{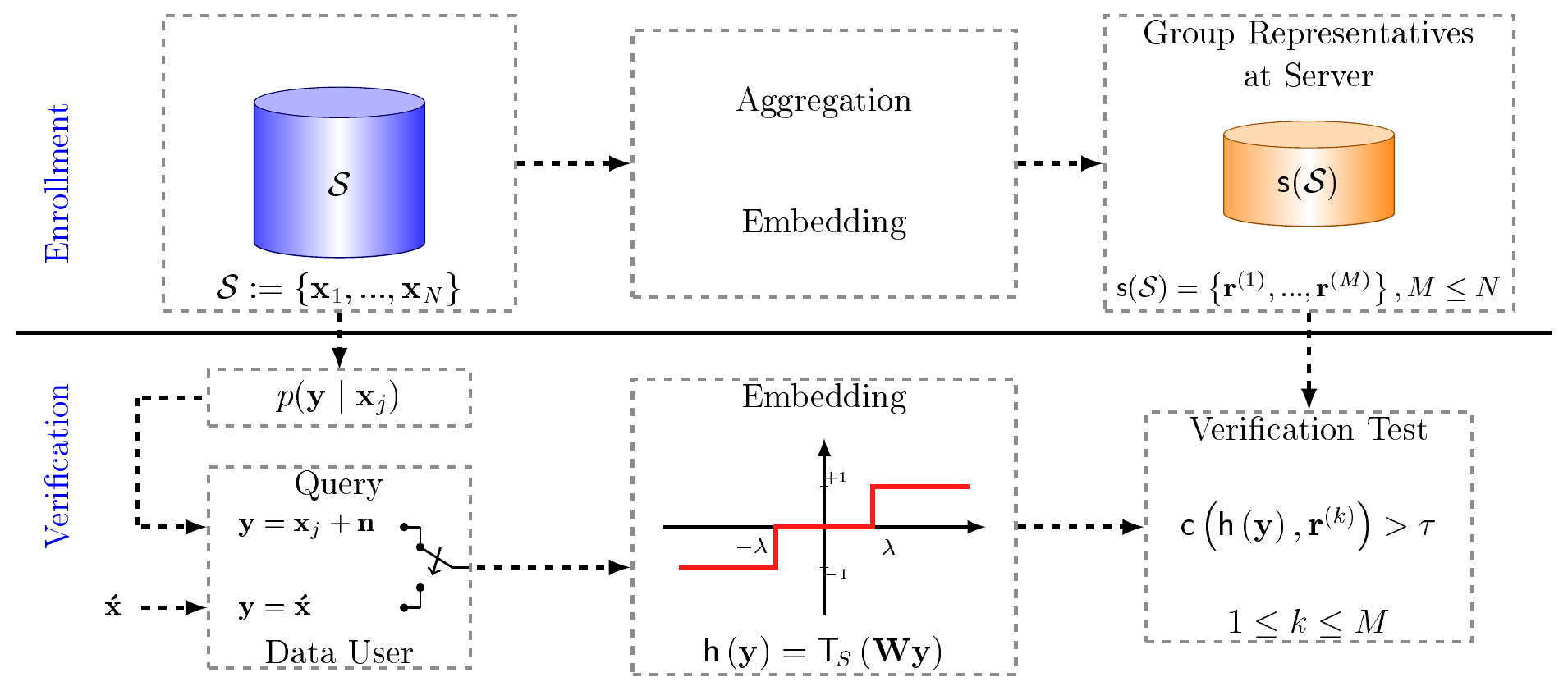}
\vspace{-8pt}
\caption{Block diagram of the proposed model.}
\vspace{-10pt}
\end{figure}

This paper proposes a group membership verification protocol preventing a curious but honest server from reconstructing the enrolled signatures and inferring the identity of querying (trusted) clients. It combines two building blocks:

\noindent \textbf{Block \#1:} One building block hashes continuous vectors into discrete \emph{embeddings}. This lossy process limits the ability of the server to reconstruct signatures from the embeddings.
\vspace{-2pt}
\noindent \textbf{Block \#2:} The other building block \emph{aggregates} multiple vectors into a unique representative value which will be enrolled at the server. The server can therefore not infer any specific signature from this value. Sufficient information must be preserved through the aggregation process for the server to assert whether or not a querying signature is a member of the group.

These two blocks can be assembled according to two configurations: block~\#1 before block~\#2, the system acquires and then hashes the signatures before aggregating them.
The opposite configuration is where acquired signatures are aggregated before hashing the result of this aggregation.
At query time, the newly acquired signature is always hashed before being sent to the server.
Weaknesses and strengths of these two configurations are explored in the paper.


%% file: RelatedWorks.tex
\vspace{-8pt}
\section{Related Work}
\label{sec:RelatedWorks}
\vspace{-8pt}

Group membership verification protocols relying on cryptography exist~\cite{Schechter:1999qy} but are more relevant to authentication, identification and secret binding. Other approaches apply homomorphic encryption to signatures, compare~\cite{Troncoso-Pastoriza:2013hi} and threshold them~\cite{Erkin:2009cz,Sadeghi:2010bl} in the encrypted domain, and need active participation of clients. Approaches involving cryptography, however, are extremely costly, memory and CPU wise.

Group membership is linked to Bloom filters that are used to test whether an element is a member of a set.
When considering security, a server using Bloom filters cannot infer any information on one specific entry~\cite{Bianchi:2012sy}.
Note that Bloom filters can not deal with continuous high dimensional signatures and that queries must be encrypted to protect the privacy of users~\cite{Boneh:2007om,Beck:2013df}. Bloom filters, adapted to our setup, however, form a baseline in our experiments (see Sect.~\ref{sec:VerifPerf}).

Embedding \emph{a single} high dimensional signature is quite a standard technique. The closest to our work is the privacy-preserving identification mechanism based on sparsifying transform~\cite{Razeghi2017wifs,Razeghi2018icassp,Razeghi2018eusipco}. It produces an information-preserving sparse ternary embedding, ensuring privacy of the data users and security of the signature.

Aggregating signals into similarity-preserving representations is very common in computer vision~\cite{Sivic:2003qp,jegou:inria-00633013,Perronnin:2007qm}. They do not consider security or privacy.
In~\cite{iscen:hal-01481220}, Iscen \etal\ use the \emph{group testing} paradigm to pack a random set of image signatures into a unique high-dimensional vector. It is therefore an excellent basis for the aggregation block: the similarities between the original non-aggregated signatures and a query signature is preserved through the aggregation.


%% file: ProblemFormulation.tex
\vspace{-8pt}

\section{Notations and definitions}
\vspace{-8pt}
Signatures are vectors in $\real^{\dim}$. If $N$ users/items belong to the group, then the protocol considers $N$ signatures, $\group=\{\x_{1}$, \ldots, $\x_{N}\}\subset \real^{\dim}$. The signature to verify is a query vector $\y\in\real^{\dim}$.  Group membership verification considers two hypotheses linked to the continuous nature of the signatures:

$\hyp_{1}$: The query is related to one of the $N$ vectors.
For instance, it is a noisy version of vector $j$, $\y=\x_{j}+\n$, with $\n$ to be a noise vector.
 
$\hyp_{0}$: The query is not related to any vector in the group.


\vspace{4pt}
We first design a group aggregation technique $\func{s}$ which computes a single representation from all $N$ vectors $\rep:= \func{s}(\group)$. This is done at the enrollment phase. Variable $\ell$ denotes the size in bits of this representation. 

At the verification phase, the query $\y$ is hashed by a function $\func{h}$ of size $\ell$ in bits.  This function might be probabilistic to ensure privacy.  The group membership test decides which hypothesis is deemed true by comparing $\func{h}(\y)$ and~$\rep$.  This is done by first computing a score function $\func{c}$ and thresholding its results: $t:=[\func{c}(\func{h}(\y),\rep)>\tau]$.

\vspace{-9pt}

\subsection{Verification Performances}
\vspace{-5pt}

The performances of this test are measured by the probabilities of false negative, $\pfn(\tau):=\Pr(t=0|\hyp_{1})$, and false positive, $\pfp(\tau):= \Pr(t=1|\hyp_{0})$.
As $\tau$ varies from $-\infty$ and $+\infty$, these measures are summarized by the AUC (Area Under Curve) performance score.
Another figure of merit is $\pfn(\tau)$ for $\tau$ s.t. $\pfp(\tau)=\epsilon$,  a required false positive level.

\vspace{-9pt}

\subsection{Security and Privacy}
\vspace{-5pt}
A curious server can reconstruct a signature $\x$ from its embedding (for instance the query): $\hat{\x} = \func{rec}(\func{h}(\x))$.
The mean squared error is a way to assess its accuracy: 
$\func{MSE} = \E(\|\X -\func{rec}(\func{h}(\X))\|^{2})/d.$
The best reconstruction is known to be the conditional expectation: $\hat{\x} = \E(\X|\func{h}(\x))$.

Reconstructing an enrolled signature from the group representation is even more challenging. 
Due to the \emph{aggregation} block, 
the curious server can only reconstruct a single vector $\hat{\x}$ from the aggregated representation, and this vector serves as an estimation of any signatures in the group:
\vspace{-8pt}
\begin{equation}
\label{eq:MSE_e}
\func{MSE}_{e} = (dN)^{-1}\sum_{j=1}^{N}\E(\|\X_{j}  -\hat{\X}\|^{2}).
\end{equation}

\vspace{-8pt}

%% file: Figure12.tex
\begin{figure}[tbp]
\begin{center}
\includegraphics[width=0.9\columnwidth,trim=0mm 1mm 0mm 1mm, clip=false,height=55mm]{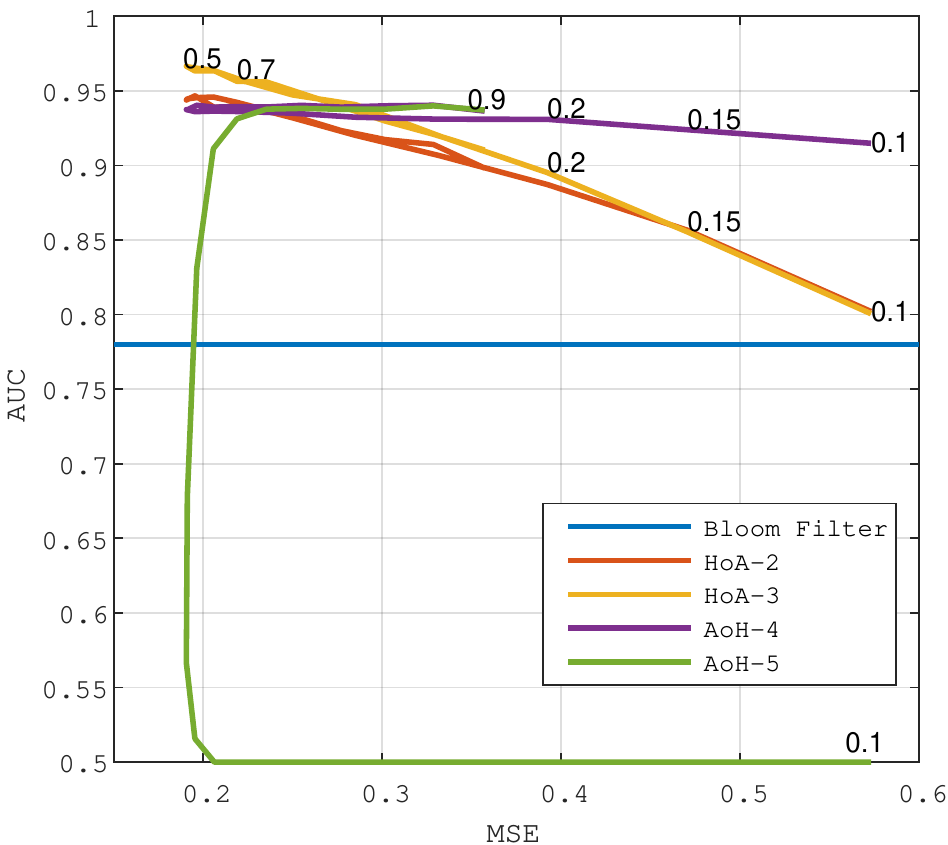}
		\caption{Unique group: $\func{AUC}$ vs. $\func{MSE}_{q}/\sigma_{y}^{2}$. $N = 128$, $\dim=1024$ , $\sigma_{n}^{2} = 0.01$ for varying $S\in (0.1\times d,0.9\times d)$.}
		\label{fig:compare}
\end{center}
\vspace{-0.7cm}
\end{figure}

\begin{figure}[bp]
\begin{center}
\includegraphics[width=0.9\columnwidth,trim=0mm -1.5mm 0mm 6mm,clip=false,height=52mm]
		{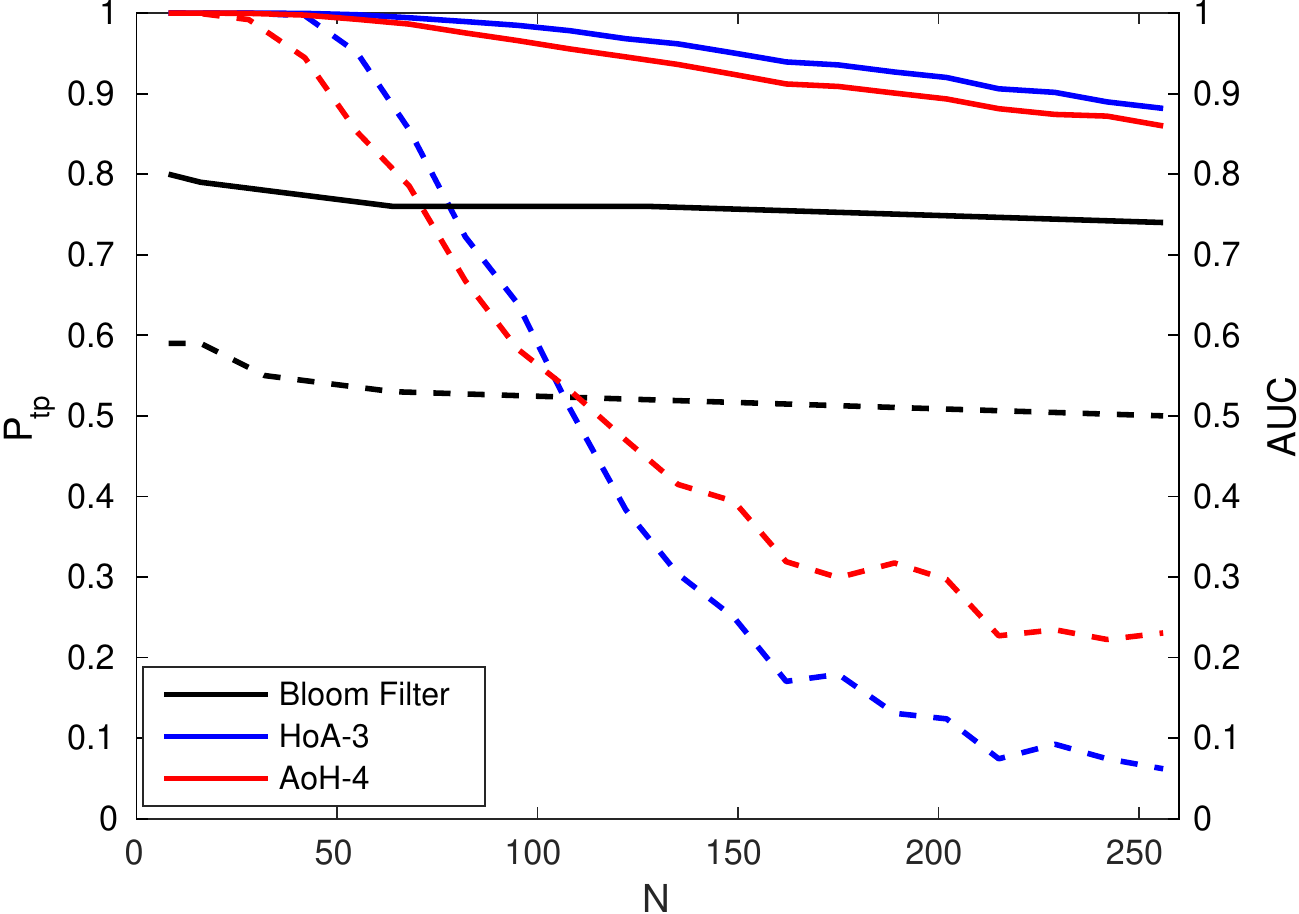}
		\caption{Unique group: $\func{AUC}$ and $\ptp$ vs. $N$. Solid and dashed lines correspond to  $\func{AUC}$ and $\ptp @ \pfp=10^{-2}$.}
		\label{fig:AUCN}
\end{center}
\vspace{-0.7cm}
\end{figure}

%% file: OneGroup_L.tex
\vspace{-8pt}

\section{Verification for a few group members} 
\label{sec:OneGroup}
\vspace{-4pt}
This section discusses the verification protocol
when $N$ is small.
We study the two different configurations for assembling block~\#1 and block~\#2. 

\textbf{Block \#1: Embedding. }
An embedding $\func{h}:\real^{\dim}\to \mathcal{A}^{\ell}$ maps a vector to a sequence of $\ell$ discrete symbols.
This quantization shall preserve enough information to tell whether two embeddings are related, but not enough to reconstruct a signature.
We use the sparsifying transform coding~\cite{Razeghi2017wifs, Razeghi2018icassp}. 
It projects $\mathbf{x} \in \mathbb{R}^{d}$ to the range space of a transform matrix $\mathbf{W} \in \mathbb{R}^{\ell \times d}$.
The output alphabet $\mathcal{A}=\{-1, 0, +1 \}$ is imposed by quantizing the components of $\mathbf{W}\mathbf{x}$ whose amplitude is lower than $\lambda$ to 0, the others to +1 or -1 according to their sign.
In expectation, $S = 2d\Phi(-\lambda/\sigma_{x})$ symbols are  non null.



\textbf{Block \#2: Aggregation.} Aggregation $\func{a}$ 
processes a set of input vectors to produce a unique output vector. When block \#1 is used before, 
that is, when considering $\func{s} = \func{a} \circ \func{h}$ , then $\func{a}:\real^{\ell\times N}\rightarrow \real^{\ell}$. When block \#2 is used before block \#1, that is when considering $\func{s} = \func{h} \circ \func{a}$, then $\func{a}:\real^{\dim\times N}\rightarrow \real^{\dim}$.

\vspace{-7pt}
\subsection{Aggregation strategies}
\vspace{-4pt}
\label{sec:AggStrat}
The nature of $\func{a}$ highly depends on the type of vector the aggregation function receives. When considering $\func{s} = \func{h} \circ \func{a}$, then $\func{a}$ gets continuous signatures. In this case it is possible to design two aggregations schemes that are: \vspace{-4pt}
\begin{eqnarray}
\func{a}(\group)&=&\sum_{\x\in\group}\x = \G \mathbf{1}_N \quad\text{or}\label{eq:SumAgg}\\
\func{a}(\group)&=&(\G^\dagger)^\top\mathbf{1}_N.\label{eq:PinvAgg}
\end{eqnarray}
where $\G$ is the $\dim\times N$ matrix $\G:=[\x_{1},\ldots,\x_{N}]$, $\mathbf{1}_N:=(1,\ldots,1)^{\top}\in\real^{N}$, and $\G^{\dagger}$ is the pseudo-inverse of $\G$. Eq.~\eqref{eq:SumAgg} is called the sum and~\eqref{eq:PinvAgg} the pinv schemes in~\cite{iscen:hal-01481220}. 

When considering $\func{s} = \func{a} \circ \func{h}$, then $\func{a}$ gets the embeddings of the signatures.
Two additional aggregation strategies are
the sum and sign pooling~\eqref{eq:AggSumSign} and the majority vote~\eqref{eq:AggMaj}:\vspace{-3pt}
\begin{eqnarray}
\rep&=&\func{sign}(\sum_{\x\in\group}\func{h}(\x))\quad\text{or}\label{eq:AggSumSign}\\
r_{i}&=&\arg\max_{s\in\{-1,0,1\}} |\{\x\in\group|\func{h}(\x)_{i} = s\}|\label{eq:AggMaj}
\end{eqnarray}

\vspace{-16pt}
\subsection{Four resulting schemes}
\vspace{-4pt}
The assemblage of the blocks and the aggregation strategies overall create four variants.
We name them:
\begin{itemize}
\item \textbf{HoA-\ref{eq:SumAgg}:} this scheme sums the raw signatures into a unique vector before embedding it in order to obtain $\rep$. It therefore corresponds to the case where $\func{s} = \func{h} \circ \func{a}$, the aggregation $\func{a}$ being defined by~\eqref{eq:SumAgg}. 
\item \textbf{HoA-\ref{eq:PinvAgg}:} here also, aggregation precedes embedding,
$\func{s} = \func{h} \circ \func{a}$, and $\func{a}$ is defined by~\eqref{eq:PinvAgg}. 
\item \textbf{AoH-\ref{eq:AggSumSign}:} this scheme embeds each signature before aggregating
with sum and sign pooling
as defined by~\eqref{eq:AggSumSign}. 
\item \textbf{AoH-\ref{eq:AggMaj}:} here also, embedding precedes aggregation, but the majority vote is used
as defined by~\eqref{eq:AggMaj}. 
\end{itemize}

The score function $\func{c}$ comparing the hashed query with the group representation is always $\func{c}(\func{h}(\y),\rep) = -\|\func{h}(\y) - \rep\|$.

\vspace{-6pt}
\section{Reconstruction and Verification}
\vspace{-4pt}
This section makes the following assumptions: i) Enrolled signatures are modelled by $\X\sim\mathcal{N}(\0_{d},\sigma_{x}^{2}\mathbf{I}_{d})$, ii) Square orthogonal matrix $\W$ known by the attacker.

\vspace{-6pt}
\subsection{Ability to reconstruct from the embedding}
\vspace{-4pt}
Now that $\W$ preserves the norm, the $\func{MSE}$ on $\X$ is the same as the mean square reconstruction error on $\Z = \W\X$, which is also white Gaussian distributed. Thanks to the independance of the components of $\Z$, the conditional expectation can be computed component-wise.
We introduce the density function conditioned on the interval $\mathcal{R}_{s}\subset\real$:\vspace{-4pt}
\begin{equation}
f(z|\mathcal{R}_{s}):= \phi_{\sigma_{x}}(z).\un_{\mathcal{R}_{s}}(z) / \Pr(Z\in\mathcal{R}_{s}),
\end{equation}
with intervals $\mathcal{R}_{0} = [-\lambda,\lambda]$, $\mathcal{R}_{1} = (\lambda,+\infty)$, and $\mathcal{R}_{-1} = (-\infty,-\lambda)$.
Function $\phi_{\sigma_{x}}$ is the p.d.f. of $Z\sim\mathcal{N}(0;\sigma_{x}^{2})$ and $\un_{\mathcal{R}_{s}}$ is the indicator function of interval $\mathcal{R}_{s}$.

Observing the $i$-th symbol of $\func{h}(\x)$ equals $s$ reveals that $z_{i}\in\mathcal{R}_{s}$.
This component is reconstructed as $\hat{z}_{i}(s) := \E(Z|\mathcal{R}_{s})$.
Note that $\hat{z}_{i}(0)=0$ because $f(z|\mathcal{R}_{0})$ is symmetric around 0.
For $s=1$, the reconstruction value equals
$\hat{z}_{i}(1)=\int_{-\infty}^{+\infty} z.f(z|\mathcal{R}_{1})dz= \frac{\sigma_{y}}{p_{1}\sqrt{2\pi}}e^{-\frac{\lambda^{2}}{2\sigma_{x}^{2}}}$
, where $p_{1}:=\Pr(Z\in\mathcal{R}_{1}) = \Phi(-\lambda/\sigma_{x})$. 
By symmetry, 
$\hat{z}_{i}(-1) = -\hat{z}_{i}(1)$, and $\func{MSE}$ admits the following close form:
\begin{eqnarray}
\func{MSE}&=&\sigma_{x}^{2}.\func{MSE}(\lambda)\label{eq:MSEb}\\
\func{MSE}(\lambda)&:=&1 - \frac{1}{\pi\Phi(-\lambda/\sigma_{x})}e^{-\frac{\lambda^{2}}{\sigma_{x}^{2}}}.
\end{eqnarray}
This quantity starts at $1-2\pi^{-1}$ when $\lambda = 0$. The embeddings are then full binary words ($p_{1}=1/2$).
All components are reconstructed by $\pm \hat{z}_{i}$ but with a large variance. As $\lambda$ increases, this variance decreases but less non-null components are reconstructed. $\func{MSE}(\lambda)$ achieves a minimum of $\approx 0.19$ for $\lambda\approx 0.60$,
where $55\%$ of the symbols of an embedding are non null.
Then, $\func{MSE}(\lambda)$ increases up to $1$ for a large $\lambda$: the embeddings becomes sparser and sparser.
When fully zero, each component is reconstructed by $0$, and $\func{MSE}$ equals $\sigma_{x}^{2}$.

\input{Figure34}

\vspace{-6pt}

\subsection{Ability to reconstruct the signatures}
\vspace{-4pt}
The curious server tries to reconstruct a unique vector $\hat{\x}$ from $\rep$ which represents the $N$ enrolled signatures. Note that $\rep$  is scale invariant: scaling the signatures by any positive factor does not change $\rep$.
Suppose that the curious server reconstructs $\hat{\x}=\kappa\w$. The best scaling minimizing $\func{MSE}_{e}$~\eqref{eq:MSE_e} is:\vspace{-1pt}
$\kappa^{\star} = \|\w\|^{-2}\w^{\top} \m,$
with $\m:= N^{-1} \sum_{j=1}^{N}\x_{j}$.
The curious server can not compute $\kappa^{\star}$ giving birth to a larger distortion:
\begin{equation}
\func{MSE}_{e} \geq \frac{1}{N} \sum_{j=1}^{N}\|\x_{j}\|^{2} - \frac{(\w^{\top}\m)^{2}}{\|\w\|^{2}}.
\end{equation}
This lower bound is further minimized by choosing $\w\propto \m$.

Therefore, aggregation~\eqref{eq:SumAgg} is less secure as the other schemes do not allow the reconstruction of $\m$.
In the worst case~\eqref{eq:SumAgg}, the curious server estimates $\m$ by 
$N^{-1}\func{rec}(\func{h}(\func{a}(\group)))$:
\begin{eqnarray}
d.\func{MSE}_{e}&\hspace{-0.35cm}=&\hspace{-0.35cm}\E(\|\X_{j} - N^{-1} \func{rec}(\func{h}(\func{a}(\group)))\|^{2})\\
&\hspace{-0.35cm}=&\hspace{-0.35cm} \E(\|\X_{j} - \frac{\func{a}(\group)}{N}\|^{2}) + \frac{\E(\|\func{a}(\group)-\func{rec}(\func{h}(\func{a}(\group)))\|^{2})}{N^{2}}.\nonumber
\end{eqnarray}

The first term is the squared distance between $\X_{j}$ and $\m$, whereas the second term corresponds to the error reconstruction for inverting the embedding. In the end:\vspace{-4pt}
\begin{equation}
\func{MSE}_{e}= \sigma_{x}^{2} \left(1-\frac{1}{N}(1-\func{MSE}(\lambda))\right).
\end{equation}
This figure of merit increases with $N$ because $\func{MSE}(\lambda)\leq 1$, $\forall \lambda\geq0$: Packing more signatures increases security.

\vspace{-6pt}
\subsection{Verification performances}
\label{sec:VerifPerf}
\vspace{-4pt}
We compare to a baseline defined as a Bloom filter optimally tuned for given $N$ and $\pfp$ having length $\ell_{B}=\lceil N|\log \pfp|\log(2)^{-2} \rceil$. An embedding $\func{h}$ is mandatory to first turn the real signatures into discrete objects.
This means that, under $\hyp_{1}$, a false negative happens whenever $\func{h}(\x_{j}+\n)\neq \func{h}(\x_{j})$.  

Fig.~\ref{fig:compare} shows the  $\func{AUC}$ vs. $\func{MSE}$~\eqref{eq:MSEb} for the schemes of Sect.~\ref{sec:AggStrat} for different sparsity $S/d$. Two schemes performs better. For low privacy (small $\func{MSE}_{q}$), HoA-\ref{eq:PinvAgg} achieves the largest $\func{AUC}$ (with $0.5\leq S/\dim \leq 0.7$)~; for high privacy, AoH-\ref{eq:AggSumSign} is recommended (with $S/\dim \leq 0.2$). In these regimes, the performances are better than the Bloom filter. 

Fig.~\ref{fig:AUCN} shows how the verification performances decrease as the number $N$ of enrolled signatures increases.
As mentioned in~\cite{iscen:hal-01481220}, the behavior of the aggregation scheme depends on the ratio $N/\dim$.
The longer the signatures, the more of them can be packed into one representation.

%% file: Figure34.tex

	\begin{figure}[tbp]
\begin{center}
\includegraphics[width=0.45\textwidth,trim=0mm 0mm 0mm 0mm, clip=false,height=45mm]{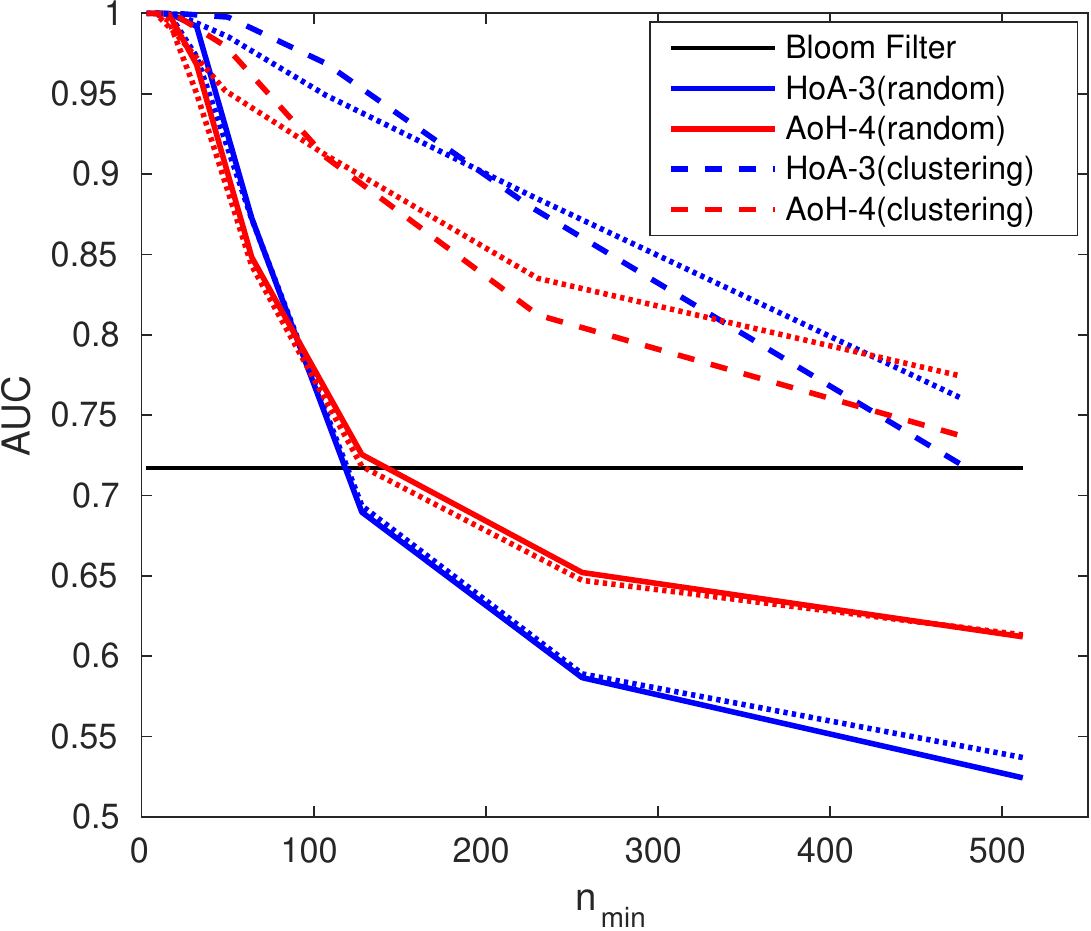}
\caption{Multiple groups: $\func{AUC}$ vs. $n_{\min}$. Dotted lines are theoretical $\func{AUC}$. $N = 4096$, $\dim=1024$, $\sigma_{n}^{2} = 10^{-2}$, $S/\dim=0.6$ for HoA-3, and $S/\dim = 0.85$ for AoH-4.}
		\label{fig:Theory}
\end{center}
\vspace{-0.7cm}
\end{figure}

\begin{figure}[tbp]
\begin{center}
\includegraphics[width=0.45\textwidth,trim=0mm 0mm 0mm 0mm,clip=false,height=45mm]
		{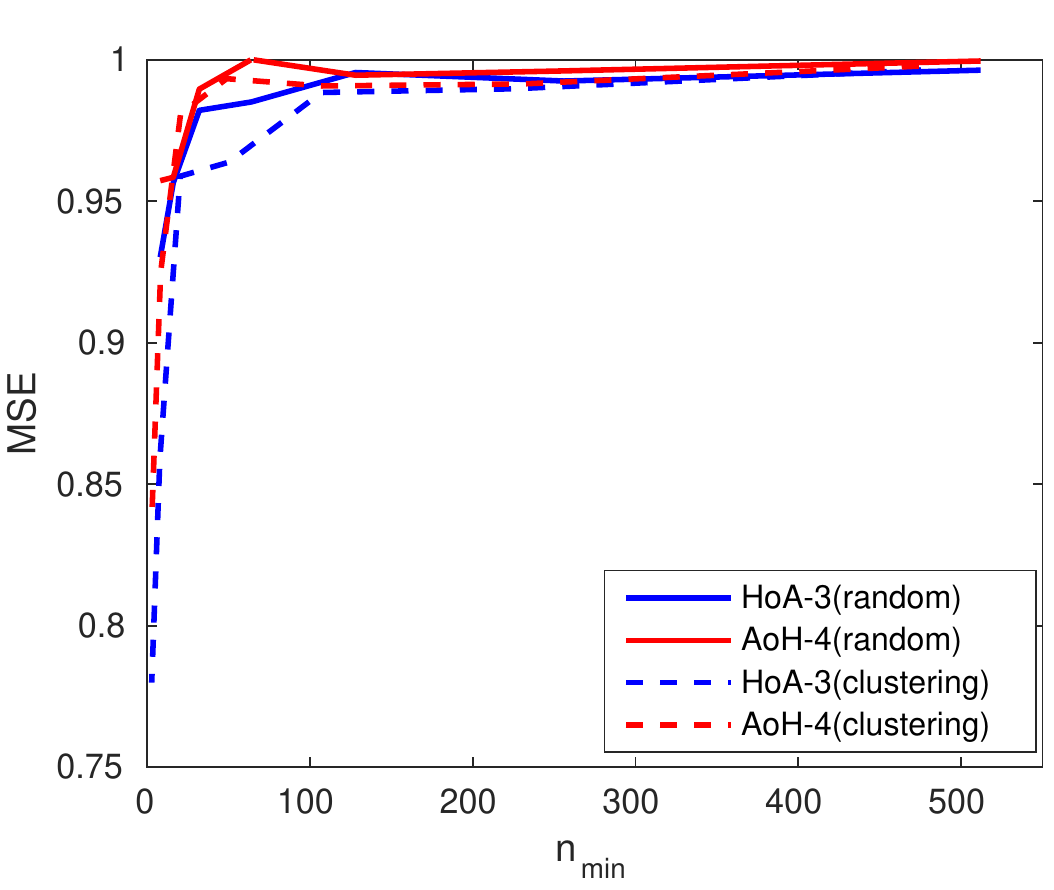}
		\caption{Multiple groups: $\func{MSE}_{e}$ vs. $n_{min}$. $N = 4096$, $\dim=1024$, $\sigma_{n}^{2} = 10^{-2}$, $S/\dim=0.6$ (HoA-3) or $0.85$ (AoH-4).}
		\label{fig:MSEM}
\end{center}
\vspace{-0.7cm}
\end{figure}

%% file: MGroups.tex
\vspace{-8pt}
\section{Verification for multiple groups}
\label{sec:MGroups}
\vspace{-4pt}
When $N$ is large, aggregating all the signatures into a unique $\rep$ performs poorly. Rather, for large $N$, we propose to partition the enrolled signature into $M>1$ groups, and to compute $M$ different representatives, one per partition.   

\noindent \textbf{Random assignment:} The signatures are randomly assigned into $M$ groups of size $n=N/M$.

\noindent \textbf{Clustering:} Similar signatures are assigned to the same group.
	The paper uses the k-means algorithm to do so. Yet, the size of the groups is no longer constant.
	

\vspace{-9pt}
\subsection{Verification performances}

\vspace{-8pt}
Denote by $(\pfp^{(k)},\ptp^{(k)})$ the operating point of group number $k$, $1\leq k\leq M$.
The overall system outputs a positive answer when at least one group test is positive.
Denote by $(\Pfp(M),\Ptp(M))$ the performance of the global system.
Under $\hyp_{0}$, the query is not related to any vector. Therefore, \vspace{-6pt}
\begin{equation}
\label{eq:PfpMgroups}
\Pfp(M) = 1 -\prod_{k=1}^{M} (1 - \pfp^{(k)}),
\end{equation}
Under $\hyp_{1}$, the query is related to only one vector belonging to one group.
A false negative occurs, if this test produces a false negative and the other tests a true negative each:\vspace{-7pt}
\begin{equation}
\label{eq:PfnMgroups}
\Pfn(M) =\sum_{k=1}^{M} \frac{n_k}{N}\pfn^{(k)}\prod_{l\neq k}(1-\pfp^{(l)}).
\end{equation}
The operating point of a group test is mainly due to the size of the group.
The random assignment creates even groups (if $M$ divides $N$),
so these share the operating point $(\pfp,\ptp)$.



Fig.~\ref{fig:Theory} shows the experimental $\func{AUC}$ and the one predicted by~\eqref{eq:PfpMgroups} and~\eqref{eq:PfnMgroups} when $M$ ranges from $8$ to $512$.
Since clustering makes groups of different sizes, we show the performances versus $n_{min}=\min_{1\leq k\leq M}(n_k)$, where $n_k$ is the size of $k$-th group.
The theoretical formulas are more accurate for random partitioning where the group are even. 
Estimations of $(\pfp^{(k)},\pfn^{(k)})$ were less precise with the clustering strategy, and this inaccuracy cumulates in~\eqref{eq:PfpMgroups} and~\eqref{eq:PfnMgroups}.  

Clustering improves the verification performances a lot especially for HoA-\ref{eq:PinvAgg}.
A similar phenomenon was observed in~\cite{iscen:hal-01481220}. 
Yet, Fig.~\ref{fig:MSEM} shows that it does not endanger the system:
$\func{MSE}_{e}$ is only slightly smaller than for random assignment, and indeed close to 1 for $n_{\min}\geq 100$.
This is obtained for $M=32$ for HoA-\ref{eq:PinvAgg} giving $\func{AUC} = 0.97$.
The space is so big that the clusters are gigantic and not revealing much about where the signatures are. 
However, the anonymity is reduced because the server learns which group provided a positive test.
This is measured in term of $k$-anonymity by the size of the smallest group, \ie\ $n_{\min}$.
Fig.~\ref{fig:Theory} indeed shows the trade-off between $k$-anonymity and the verification performances.



